\documentclass[aps,prl,twocolumn,floatfix,groupedaddress,footinbi,bnotitlepage,showpacs]{revtex4-1}
\usepackage{graphicx,graphics,times,bm,bbm,bbold,amssymb,amsmath,amsfonts,dsfont,hyperref}

\newcommand{\ket}[1]{|#1\rangle}

\newcommand{\braket}[2]{\langle #1|#2\rangle}
\newcommand{\ketbra}[2]{|#1\rangle\langle #2|}
\newcommand{\mrm}[1]{\mathrm{#1}}
\newcommand{\ave}[1]{\langle{#1}\rangle}

\let\oldsqrt\sqrt
\def\sqrt{\mathpalette\DHLhksqrt}
\def\DHLhksqrt#1#2{%
\setbox0=\hbox{$#1\oldsqrt{#2\,}$}\dimen0=\ht0
\advance\dimen0-0.2\ht0
\setbox2=\hbox{\vrule height\ht0 depth -\dimen0}%
{\box0\lower0.4pt\box2}}

\DeclareFontFamily{OT1}{pzc}{}
\DeclareFontShape{OT1}{pzc}{m}{it}%
              {<-> s * [1.25] pzcmi7t}{}
\DeclareMathAlphabet{\mathpzc}{OT1}{pzc}%
                                 {m}{it}
\begin{document}

\title{Quantum Metrology: Extended Convexity of Quantum Fisher Information}
\author{S. Alipour}
\affiliation{Department of Physics, Sharif University of Technology, Tehran 14588-89694, Iran}
\author{A.  T. Rezakhani}
\affiliation{Department of Physics, Sharif University of Technology, Tehran 14588-89694, Iran}

\begin{abstract}
We prove an extended convexity for quantum Fisher information of a mixed state with a given convex decomposition. This convexity introduces a bound which has two parts: i. \textit{classical} part associated to the Fisher information of the probability distribution of the states contributing to the decomposition, and ii. \textit{quantum} part given by the average quantum Fisher information of the states in this decomposition. Next we use a non-Hermitian extension of symmetric logarithmic derivative in order to obtain another upper bound on quantum Fisher information, which enables to derive a closed form for a fairly general class of system dynamics given by a dynamical semigroup. We combine our two upper bounds together in a general (open system) metrology framework where the dynamics is described by a quantum channel, and derive the ultimate precision limit for quantum metrology. We illustrate our results and their applications through two examples, where we also demonstrate that how the extended convexity allows to track transition between quantum and classical behaviors for an estimation precision.
\end{abstract}

\maketitle
%%%%%%%%%%%%%%%%%%%%%%%%%%%%%%%%%%%%%%%%%%%%%%%%%%%%%%%%%%%%%%%%%%%

\textit{Introduction.---}Advent of ultra-precise quantum technologies in recent years has spurred the need for devising metrological protocols with the highest sensitivity allowed by laws of physics. Quantum metrology \cite{Lloyd:NP} investigates fundamental limits on the estimation error through the quantum Cr\'amer-Rao bound \cite{Helstrom:book,book:Holevo,Braunstein-Caves:QFI}. Without using quantum resources, the very central limit theorem indicates that parameter estimation error is bounded by the ``shot-noise limit" \cite{Estimation:book}; however, employing quantum resources, such as quantum correlations between probes \cite{qcorrelation,BAR:dissipative-manybody,Marzolino}, allows for scaling of the error to beat the shot-noise limit and reach the more favorable ``Heisenberg (or sub-shot-noise) limit" \cite{Heisenberg}---and perhaps beyond \cite{Boixo-etal:PRL07}. This feature of quantum metrology has been realized experimentally \cite{experiments}.

In realistic systems, interaction with environment is inevitable. Since quantum procedures are susceptible to noise, formulation of a framework for noisy/dissipative quantum metrology is required \cite{Lloyd}. Recently, some attempts have been made toward proposing systematic analysis of open-system quantum metrology \cite{OSQM}, where some purification for density matrices has been used \cite{Tsang,Escher:NatPhys,AMR:dissipative CRB}. Some other methods based on different approaches such as using right/left logarithmic derivative  \cite{Fujiwara-Imai} and the channel extension idea \cite{Kolodynski,Guta:NatureC} have also been proposed.

Exact calculation of quantum Fisher information (QFI) in general is difficult since it needs diagonalization of the system density matrix, which appears through the key quantity of symmetric logarithmic derivative (SLD). Besides, it is also not straightforward (except when the dynamics is unitary) to recognize from the exact form of QFI the role of underlying physical properties of the system of interest in when scaling of the estimation error behaves classically or quantum mechanically. Given these difficulties, resorting to upper bounds on QFI can be beneficial both theoretically and practically \cite{Escher:NatPhys,Kolodynski,Guta:NatureC,AMR:dissipative CRB,BAR:dissipative-manybody}. In deriving such bounds, different properties of QFI may prove useful. Convexity is an appealing property, which unfortunately does not hold for QFI in general \cite{Yu,Toth}. Notwithstanding, here we derive an \textit{extended} convexity relation for QFI, which obviously gives rise to an upper bound on QFI. We remind that every quantum state can be written (in infinite ways) as a convex decomposition of states which prepare the very state when mixed according to a given probability distribution. Having such a decomposition, we show that the upper bound contains ``classical" and ``quantum" parts. The classical part is the Fisher information associated to the (classical) probability distribution of the mixture, and the quantum part is related to the weighted average of the QFI of the constituting states of the mixture. This result is completely general and always holds---unlike some earlier results in the literature \cite{Guta:NatureC,Yu}. Additionally, we show that how having such a classical-quantum picture for QFI enables us to find when a quantum metrology scenario exhibits either of classical (shot-noise) or quantum (Heisenberg) regimes.

We also employ an extension of SLD which is non-Hermitian (hereafter, nSLD), and define an extended QFI which is shown to upperbound QFI. This nSLD has this extra utility that for dynamics with a semigroup property \cite{ }, its associated (extended) QFI is directly related to the quantum jump operators of the dynamics. In addition, we show that this extended QFI (irrespective of the underlying dynamics) for a density matrix is the same as the Uhlmann metric, obtained earlier in the context of geometry of a state \cite{Uhlmann:PT,uniform2}. This endows a geometric picture to nSLD. Our nSLD concept also allows to supplement the extended convexity property for the case of a general open quantum dynamics given by a quantum channel. Interestingly, by putting the concepts of the extended convexity and nSLD together, we recover the exact QFI for an open system \cite{Escher:NatPhys}, whence the ultimate precision for estimation of a parameter of an open system. We illustrate utility of our results through two important examples.

%%%%%%%%%%%%%%%%%%%%%%%%%%%%%%%%%%%%%%%%%%%%%%%%%%%%%%%%%%%%%%%%%%%
\textit{Extended convexity of QFI.---}We first briefly remind Fisher information and its role in metrology. In estimation of a parameter $\mathrm{x}$ of a classical system, the estimation error $\delta \mathrm{x}$ is lowerbounded by the inverse square of the classical Fisher information \cite{Cramer:book}
\begin{equation}
\mathpzc{F}^{(\mathrm{C})}_{\mathrm{x}}(\{p\})=\int_{\mathpzc{D}_{\mathrm{x}'}} d\mathrm{x}' \big(\partial_{\mrm{x}}p(\mathrm{x}'|\mathrm{x})\big)^2/{p(\mathrm{x}'|\mathrm{x})},
\label{CFI-text}
\end{equation}
where $p(\mathrm{x}'|\mathrm{x})$ is the probability distribution of obtaining value $\mathrm{x}'$ (in a measurement) given the exact value of the unknown parameter is $\mathrm{x}$, and $\mathpzc{D}_{\mathrm{x}'}$ is the domain of all admissible $\mathrm{x}'$. In quantum systems, measurements are described by a set of positive operators $\{\Pi_{\mathrm{x}'}\}$ which has the completeness property $\int_{\mathpzc{D}_{\mathrm{x'}}} \mathrm{d}\mathrm{x}'\Pi_{\mathrm{x}'}=\openone$. If $\varrho_{\tau}(\mathrm{x})$ (hereafter sometimes $\varrho$ or $\varrho(\mathrm{x})$, to lighten the notation) denotes the state of system at time $\tau$, we have $p(\mathrm{x}'|\mathrm{x})=\mathrm{Tr}[\varrho_{\tau}(\mathrm{x})\Pi_{\mathrm{x}'}]$. In quantum metrology a key quantity is SLD, which is a Hermitian operator $L_{\varrho}$ satisfying
\begin{equation}
\partial_{\mathrm{x}} \varrho=(L_{\varrho}\varrho+\varrho L_{\varrho})/2.
\label{sld}
\end{equation}
Optimizing over all set of measurements yields the QFI
\begin{equation}
\mathpzc{F}^{(\mathrm{Q})}_{\mathrm{x}}(\varrho)=\mathrm{Tr}[\varrho L_{\varrho}^2],
\label{def:qfi}
\end{equation}
whereby the quantum Cr\'amer-Rao bound $\delta \mathrm{x}\leqslant\big(\mathpzc{F}^{(\mathrm{Q})}_{\mathrm{x}}\big)^{-1/2}$ gives the achievable minimum estimation error \cite{Braunstein-Caves:QFI,Paris:QEforQT}.

A quantum state $\varrho$ can be written as a mixture (i.e., convex superposition) of some quantum states $\{\varrho_{a}\}$ with some probabilities $\{p_{a}\}$ as $\varrho=\sum_a p_{a}\varrho_{a}$. Now we show that QFI possesses an \textit{extended} convexity property as follows:\begin{equation}
\mathpzc{F}^{(\mathrm{Q})}_{\mathrm{x}}(\textstyle{\sum_a} p_a \varrho_a)\leqslant \textstyle{\sum_a} p_{a}\mathpzc{F}^{(\mathrm{Q})}_{\mathrm{x}}(\varrho_a) + \mathpzc{F}^{(\mathrm{C})}_{\mathrm{x}}(\{p_{a}\}).
\label{ave-CFI-QFI}
\end{equation}
The sketch of the proof is as follows---see Ref.~\cite{supp} for details. We start by differentiating $\varrho=\sum_a p_{a}\varrho_{a}$ where we use the following relation:
\begin{equation}
\partial_{\mathrm{x}}(p_{a}\varrho_{a}) = (\mathsf{L}_a p_{a}\varrho_{a}+p_{a}\varrho_{a}\mathsf{L}_a)/2,
\label{SLD(prho)}
\end{equation}
with $\mathsf{L}_a =L_{\varrho_{a}} + (\partial_{\mathrm{x}} p_{a})/p_a$. The rest of the proof is straightforward by following the same steps as in the derivation of QFI \cite{Paris:QEforQT}. This yields the upper bound $\textstyle{\sum_a} p_{a}\mathrm{Tr}[\varrho_{a} \mathsf{L}_a^2]$ for QFI, which reproduces the right-hand side of Eq.~(\ref{ave-CFI-QFI})---named for future reference as $\mathpzc{F}^{(\mathrm{Q})}_{\mathrm{conv}}$---after replacing $\mathsf{L}_{a}$.

Several remarks are in order here. i. From Eq.~(\ref{ave-CFI-QFI}) it should be evident that the very classical term $\mathpzc{F}^{(\mathrm{C})}_{\mathrm{x}}(\{p_a\})$ obstructs the convexity to hold in general---whence ``extended" convexity. This term, however, vanishes when the mixing probabilities $p_{a}$ do not depend on $\mathrm{x}$. Such a special case occurs when $\varrho_0$ (assumed independent of $\mathrm{x}$) evolves unitarily under an $\mathrm{x}$-dependent Hamiltonian, e.g., $\mathrm{x} H$. Here, one can see that
\begin{equation}
\mathpzc{F}^{(\mathrm{Q})}_{\mathrm{conv}}=\textstyle{\sum_a}p_{a} \mathpzc{F}^{(\mathrm{Q})}(\varrho_a)=4 \tau^2\textstyle{\sum_a}p_{a} \Delta^2_{a} H,
\end{equation}
where $\sum_a p_a\varrho_a$ is the spectral decomposition of $\varrho$, and $\Delta^2_{a} H=\langle (H-\ave{H}_{\varrho_a})^2\rangle_{\varrho_a}$ [with $\ave{\circ}_{\zeta}\equiv\mathrm{Tr}[\zeta\circ]$] is the variance/uncertainty of $H$ with respect to $\varrho_{a}$~\cite{Toth,Yu}. Another case in which the classical term vanishes is when one uses the uniform ensemble decomposition for a state---such decomposition for any state indeed always exists (irrespective of its dynamics) \cite{uniform1,uniform2}. ii. For a state whose $\mathrm{x}$-dependence is generated through a ``classically simulatable" quantum channels, i.e., $\varrho(\mathrm{x})=\mathpzc{C}_{\mathrm{x}}[\varrho_0]$, where $\mathpzc{C}_{\mathrm{x}}[\circ]=\sum_a p_a(\mathrm{x}) A_{a} \circ A_a^{\dag}$ (with some $\mathrm{x}$-independent $A_a$s), the only nonvanishing contribution to Eq.~(\ref{ave-CFI-QFI}) is classical \cite{Guta:NatureC}. Nonetheless, not all channels can be simulated classically. iii. The classical part in Eq.~(\ref{ave-CFI-QFI}) scales with the number of probes $N$ as $O(N)$ [shot-noise], but the quantum part can scale beyond as $O(N^2)$ [Heisenberg]. The extended convexity relation allows to see when the estimation error scales classically or quantum mechanically, and moreover when there is a transition in this behavior. This is a possibility mostly absent before having the extended convexity property. We discuss this later through a specific example. iv. To obtain the tightest upper bound, we should perform a minimization over all ensemble decompositions of $\varrho$, namely $\min_{\{p_{a},\varrho_{a}\}}\mathpzc{F}_{\mathrm{conv}}^{(\mathrm{Q})}$. In the following we employ the concept of nSLD and prove that this minimization indeed leads to the very exact QFI of the open system.

%%%%%%%%%%%%%%%%%%%%%%%%%%%%%%%%%%%%%%%%%%%%%%%%%%%%%%%%%%%%%%%%%%%
\textit{Upper bound on QFI using an \lowercase{n}SLD.---}As we discussed earlier, the standard framework of quantum metrology relies on the concept of SLD, the Hermitian operator $L_{\varrho}$ of Eq.~(\ref{sld}). However, non-Hermitian ``right" and ``left" logarithmic derivatives have also been used in the literature to construct upper bounds on QFI \cite{book:Holevo,book:Nagaoka}. In contrast, here we introduce a non-Hermitian \textit{symmetric} logarithmic derivative (whence nSLD), which is also used to obtain an upper bound on QFI. Our specific motivation to employ nSLD is that a non-Hermitian candidate ($D_{\varrho}$) for SLD for time as a parameter can be simply read from the Lindblad dynamical equation of an open quantum system \cite{Rivas-Huelga:book}
\begin{align}
\partial_{\tau}\varrho
&=(D_{\varrho}\varrho+\varrho D_{\varrho}^{\dag})/2,
\label{open-system}
\end{align}
where $D_{\varrho}=-2i\mrm{x}_0 H- \sum_a \mrm{x}_a (\Gamma_{a}^{\dag}\Gamma_{a}-\Gamma_{a} \varrho \Gamma_{a}^{\dag} \varrho^{-1})$, $\mathrm{x}_0H$ is the (Lamb-shift modified) system Hamiltonian, and $\{\Gamma_a\}$ are quantum jump operators encapsulating together with the time-independent parameters $\{\mathrm{x}_a\}_{a=1}^{k}$ the effect of environment. Note that we have set throughout $\hbar\equiv 1$. For simplicity we have assumed that $\varrho$ is invertible; however, this condition can be lifted \cite{solution-nSLD}, and the extension of our results is straightforward. Inspired by Eq. \eqref{open-system}, we define the nSLD $\widetilde{L}_{\varrho}$ through
\begin{equation}
\partial_\mathrm{x}\varrho=(\widetilde{L}_{\varrho}\varrho+\varrho \widetilde{L}_{\varrho}^{\dag})/2.
\label{SLD-1-text}
\end{equation}
Following the similar procedure as to obtain QFI from classical Fisher information---now with the nSLD $\widetilde{L}_{\varrho}$---yields \cite{supp}
\begin{equation}
\mathpzc{F}^{(\mathrm{Q})}_{\mathrm{x}}(\varrho) \leqslant\mathrm{Tr}[\widetilde{L}_{\varrho} \varrho \widetilde{L}_{\varrho}^{\dag}],
\label{F_u-text}
\end{equation}
where---for future reference---we call the right-hand side the ``extended QFI" $\mathpzc{F}^{(\mathrm{Q})}_{\mathrm{ext}}$.  It is evident that when $\widetilde{L}_{\varrho}$ is Hermitian, the inequality becomes equality because $\widetilde{L}_{\varrho}=L_{\varrho}$ [Eq.~(\ref{def:qfi})]. Although with an nSLD the bound might not be achievable in some cases, it yet imposes an intrinsic quantum lower bound on the estimation error, independent of measurement, which in some cases is more feasible to calculate. It is evident that the nSLD is not unique; any operator $O$ for which $O\varrho+\varrho O^{\dag}=0$ can be added to $\widetilde{L}_{\varrho}$). Thus unlike QFI (\ref{def:qfi}), which is invariant with respect to freedom in the Hermitian SLD $L_{\varrho}$, $\mathpzc{F}^{(\mathrm{Q})}_{\mathrm{ext}}$ depends on the choice of nSLD. Thus one can minimize over nSLDs in order to obtain a tighter upper bound. We show in Ref.~\cite{supp} that this minimization has indeed a natural geometric interpretation (which can relate quantum metrology to concepts such as geometric phase \cite{Uhlmann:PT}). Any state can be decomposed (or purified) as $\varrho=\textit{w}\textit{w}^{\dag}$, with $\textit{w}=\sqrt{\varrho}U$ with an arbitrary unitary $U$, which in turn induces a metric for curves of density matrices based on the scalar product $\langle \textit{w}_1,\textit{w}_2\rangle=\mathrm{Tr}[\textit{w}_2\textit{w}_1^{\dag}]$. Minimizing length of a dynamical curve $\varrho_0\mapsto\varrho(\mathrm{x})$---finding the ``geodesic"---leads to  a ``parallel transport condition." Requesting the same geodesic condition here as well implies that nSLD reduces to (the Hermitian) SLD (\ref{sld}). The utility of the extended QFI and nSLD will be illustrated below where we use them to enhance the implications of the extended convexity property for general quantum dynamical systems.

%%%%%%%%%%%%%%%%%%%%%%%%%%%%%%%%%%%%%%%%%%%%%%%%%%%%%%%%%%%%%%%%%%%
\textit{QFI for a general dynamical quantum channels}.---A quantum state $\varrho_0$ passing through a (parameter-dependent) quantum channel $\mathpzc{E}_{\mathrm{x}}$ evolves as
\begin{equation}
\varrho(\mathrm{x})=\mathpzc{E}_{\mathrm{x}}[\varrho_0]=\textstyle{\sum_a} A_{a}(\mathrm{x}) \varrho_0 A_{a}^{\dag}(\mathrm{x}),
\end{equation}
given by a set of Kraus operators $\{A_a\}$ (which are not unique) \cite{Rivas-Huelga:book}. This (trace-preserving) completely-positive map describes the general class of (open) quantum evolutions. This form immediately implies a possible ensemble decomposition with $p_{a}=\mathrm{Tr}[A_{a} \varrho_0 A_{a}^{\dag}]$ and $\varrho_{a}=A_{a} \varrho_0 A_{a}^{\dag}/\mathrm{Tr}[A_{a} \varrho_0 A_{a}^{\dag}]$. Now we apply the extended convexity bound (\ref{ave-CFI-QFI}) in which---bearing in mind Eq.~(\ref{F_u-text}) for the extended QFI---we replace on the right-hand side $\mathpzc{F}_{\mathrm{x}}^{(\mathrm{Q})}(\varrho_a)$ with $\mathpzc{F}_{\mathrm{ext},\mathrm{x}}^{(\mathrm{Q})}(\varrho_a)$. It is straightforward to see that if in the proof of Eq.~(\ref{ave-CFI-QFI}) we use an nSLD ($\widetilde{\mathsf{L}}_{a}$) rather than SLD ($\mathsf{L}_{a}$), the final result changes to $\sum_a p_a \mathrm{Tr}[\widetilde{\mathsf{L}}_a\varrho \widetilde{\mathsf{L}}_a^{\dag}]$. From
\begin{equation}
(\widetilde{\mathsf{L}}_{a} A_{a} \varrho_0 A_{a}^{\dag}+A_{a} \varrho_0A_{a}^{\dag}\widetilde{\mathsf{L}}^{\dag}_a)/2=\partial_{\mathrm{x}}(A_{a} \varrho_0A_{a}^{\dag}),
\end{equation}
one can read a compatible nSLD defined through $\widetilde{\mathsf{L}}_{a} A_{a}=2\partial_{\mathrm{x}}A_{a}+i \eta A_{a}$, in which $\eta$ is an arbitrary real constant. Hence
\begin{equation}
\mathpzc{F}^{(\mathrm{Q})}_{\mathrm{conv},\mathrm{x}}=\textstyle{\sum_a} 4(\ave{\partial_{\mathrm{x}}A_{a}^{\dag}\partial_{\mathrm{x}}A_{a}}_{\varrho_0}-i \eta\ave{ A_{a}^{\dag}\partial_{\mathrm{x}}A_{a}}_{\varrho_0})+\eta^2,\label{eq24}
\end{equation}
where we have employed the trace-preserving property $\sum_a A_{a}^{\dag}A_{a}=\openone$. Minimization of Eq.~\eqref{eq24} over $\eta$ yields $\mathpzc{F}^{(\mathrm{Q})}_{\mathrm{conv},\mathrm{x}}=4\Big(\ave{\textsf{H}_1}_{\varrho_0}-\ave{\textsf{H}_2}_{\varrho_0}^2\Big)$, where $\textsf{H}_1=\textstyle{\sum_a} \partial_{\mathrm{x}}A_{a}^{\dag}\partial_{\mathrm{x}}A_{a}$ and $\textsf{H}_2=i\textstyle{\sum_a} A_{a}^{\dag}\partial_{\mathrm{x}}A_{a}$. Additionally, we still have the freedom to minimize over all compatible convex decomposition of $\varrho$. This minimization here translates to minimization over all Kraus operators $A_{a}$ compatible with the dynamics $\mathpzc{E}_{\mathrm{x}}$. Thus the tightest bound is obtained as
\begin{align}
\mathpzc{F}^{(\mathrm{Q})}_{\mathrm{x}}(\varrho)\leqslant 4\min_{\{A_{a}\}}\Big(\ave{\textsf{H}_1}_{\varrho_0}
-\ave{\textsf{H}_2}_{\varrho_0}^2\Big).
\end{align}
In fact, this is an \textit{equality}, because this bound has been already proved in Ref.~\cite{Escher:NatPhys}---albeit through a different approach---to be the very QFI $\mathpzc{F}^{(\mathrm{Q})}_{\mathrm{x}}(\varrho)$. This fact proves in an indirect manner that our combined convexity-nSLD approach generates the exact (hence obviously achievable) QFI.

In addition to this appealing feature of our framework, in the following we discuss two other applications of our results, which further enhance the utility of the framework.

%%%%%%%%%%%%%%%%%%%%%%%%%%%%%%%%%%%%%%%%%%%%%%%%%%%%%%%%%%%%%%%%%%%
\textit{Extended convexity and quantum-to-classical transition.---}As remarked earlier, the extended convexity relation (\ref{ave-CFI-QFI}) divides the upper bound on QFI into \textit{classical} and \textit{quantum} parts. This is a physically appealing feature in that this division enables to discern how classical and quantum parts compete in QFI (whence in the metrology precision). As a result, one may infer when there can exist a threshold number of probes $N^{\star}$ below which the error can show a Heisenberg-like scaling, while above that the error eventually reduces to the shot-noise limit (as predicted in Ref.~\cite{Guta:NatureC}). Additionally, this helps determine for probe sizes lager than a specific threshold value $N^{\star}$, using quantum resources (e.g., entanglement) to enhance estimation is ineffective---since one may reach the same estimation error exactly through classical schemes with an approximately equal probe size. Another immediate special result of our extended convexity property, as discussed in a remark after Eq.~(\ref{ave-CFI-QFI}), is that for the special class of classically-simulatable quantum channels QFI is always bounded linearly, hence associated estimation errors scale classically (i.e., shot-noise limit) \cite{Guta:NatureC}. The following example illustrates further how a quantum-to-classical transition in noisy/open quantum metrology can be identified.

\begin{figure}[tp]
\includegraphics[scale=0.37]{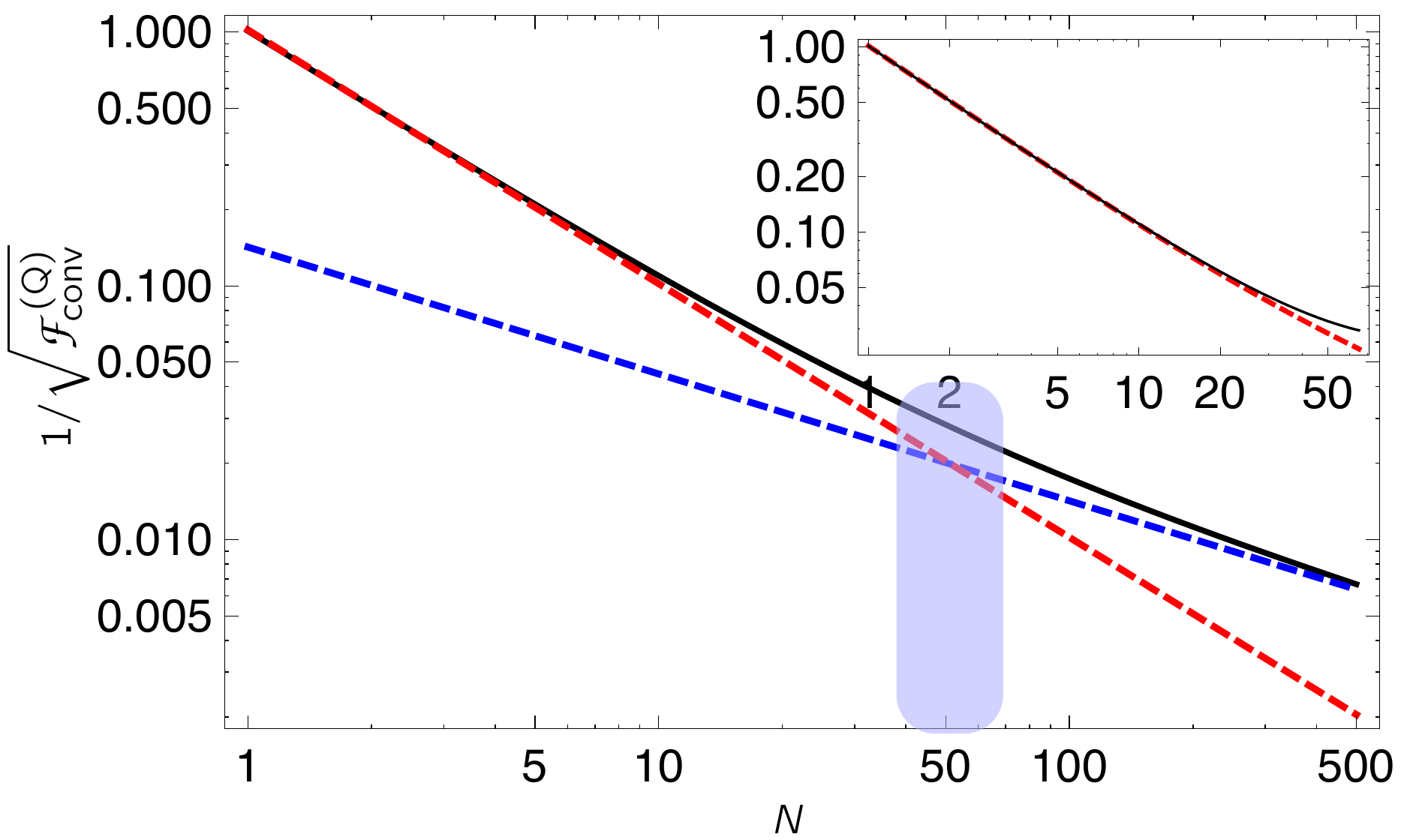}
\caption{The optimal lower bound of the error $1/\sqrt{\mathpzc{F}^{(\mathrm{Q})}_{\mathrm{conv}}}$ (solid-black) and its approximations $c_2^{-1/2}/N$  [dashed-red] and $c_1^{-1/2}/\sqrt{N}$ [dashed-blue] for $N\ll N^{\star}$ and $N\gg N^{\star}$, respectively. Here $q=0.995$, $\tau=1$, and $\mathrm{x}=1$, which yield $N^{\star}\approx 50$, agreeing also with the plot [the light blue area]. The inset compares our bound $1/\sqrt{\mathpzc{F}^{(\mathrm{Q})}_{\mathrm{conv}}}$ [dashed-red] with the exact value of $1/\sqrt{\mathpzc{F}^{(\mathrm{Q})}}$ [solid-black], which again both agree well up to $N\approx 65$.}
\label{transition}
\end{figure}

%%%%%%%%%%%%%%%%%%%%%%%%%%%%%%%%%%%%%%%%%%%%%%%%%%%%%%%%%%%%%%%%%%%
\textit{Example 1.}---Consider a dephasing quantum channel with Kraus operators $A_1=(\sqrt{q}\cos[\alpha \mathrm{x}]+i\sqrt{1-q}\sin[\alpha \mathrm{x}]\sigma_z)e^{i \mathrm{x} \tau \sigma_z/2}$ and $A_2=(i\sqrt{q}\sin[\alpha \mathrm{x}]+\sqrt{1-q}\cos[\alpha \mathrm{x}]\sigma_z)e^{i \mathrm{x} \tau \sigma_z/2}$, where $\sigma_z=\mathrm{diag}(1~-1)$ is a Pauli matrix, $\alpha \in \mathbb{R}$ is arbitrary, $0\leqslant q \leqslant 1$ characterizes the amount of loss (note that $q$ can in general depend on $\tau$), and $\mathrm{x}$ is the parameter to be estimated. If we assume $N$ initial probes each of which evolving through a separate dephasing channel, one can see that the states $\varrho_{a_1,\cdots,a_N}=A_{a_1}\cdots A_{a_N}\varrho_0 A^{\dag}_{a_1}\cdots A^{\dag}_{a_N}/p_{a_1,\cdots,a_N}$ and the probabilities $p_{a_1,\cdots,a_N}=\mathrm{Tr}[A_{a_1}\cdots A_{a_N}\varrho_0 A^{\dag}_{a_1}\cdots A^{\dag}_{a_N}]$ constitute a mixture, where $a_n\in\{1,2\}$ indicates the $n$th probe. By choosing $\varrho_0=\ketbra{\textsc{ghz}_N}{\textsc{ghz}_N}$, where $|\textsc{ghz}_N\rangle=(|0\rangle^{\otimes N}+ |1\rangle^{\otimes N})/\sqrt{2}$, Eq.~\eqref{ave-CFI-QFI} gives
%\begin{align}
%&\mathpzc{F}^{(\mathrm{C})}_{\mathrm{x}}(\{p_{a_1,\ldots,a_N}\})=\frac{4 N\alpha^2(1 - 2 q)^2\sin^2[2 \alpha \mathrm{x}]}{1-(1 - 2 q)^2\cos^2[2 \alpha \mathrm{x}]},\nonumber\\
%&\textstyle{\sum_{a_1,\ldots,a_N}} p_{a_1,\ldots,a_N}\mathpzc{F}_{\mathrm{x}}^{(\mathrm{Q})}(\varrho_{a_1,\ldots,a_N})=N^2 \big(\tau^2+\tau\alpha\nonumber\\
%&~\times\sqrt{q(1-q)}+16\alpha^2q(1-q)\big)+4N\alpha^2[1 - 4 q(1-q)]\nonumber\\
%&~-\mathpzc{F}_{\mathrm{x}}^{(\mathrm{C})}(\{p_{a_1,\ldots,a_N}\}).
%\end{align}
%Thus
$\mathpzc{F}^{(\mathrm{Q} )}_{\mathrm{conv}}=c_1(\alpha,q,\tau)N+c_2(\alpha,q,\tau)N^2$, where $c_1=4\alpha^2\big(1 - 4 q(1-q)\big))$ and $c_2=\tau^2+8\tau\alpha\sqrt{q(1-q)}+16\alpha^2q(1-q)$. From this relation, if $\alpha$ and $q$ do not depend on $N$, one can find the threshold size as $N^{\star}=c_1/c_2$. However, in finding the optimal $\mathpzc{F}^{(\mathrm{Q} )}_{\mathrm{conv}}$ (with respect to the arbitrary parameter $\alpha$), an $N$-dependent $\alpha_{\mrm{opt}}=-(N\sqrt{q(1-q)}\tau)/(1+4q(N-1)(1-q))$ arises. Choosing $N^{\star}=(c_1/c_2)|_{\alpha_{\mrm{opt}}}=(1-2q)^2/[4 q(1-q)]$ still works for this case, since it is evident that $c_1\approx (1-2p)^2 \tau^2/[16 p(1-p)]$ and $c_2\approx 0$ when $N\gg N^{\star}$, and $c_1\approx 0$ and $c_2\approx (1 - 2 p)^4 \tau^2$ when $N\ll N^{\star}$. Figure~\ref{transition} depicts this quantum-to-classical transition. Note that we have not optimized over \textit{all} compatible Kraus operators, whereby our obtained upper bound here does not necessarily follow the exact QFI. In fact, one can show that here the exact QFI vanishes exponentially with $N$ and $\tau$ \cite{Chin:PRL2012,AMR:dissipative CRB}. If we perform an exhaustive search and optimization over a larger class of Kraus operators, we expect to capture this exponential reduction of the QFI through our formalism too.
%%%%%%%%%%%%%%%%%%%%%%%%%%%%%%%%%%%%%%%%%%%%%%%%%%%%%%%%%%%%%%%%%%%
%%%%%%%%%%%%%%%%%%%%%%%%%%%%%%%%%%%%%%%%%%%%%%%%%%%%%%%%%%%%%%%%%%%

%%%%%%%%%%%%%%%%%%%%%%%%%%%%%%%%%%%%%%%%%%%%%%%%%%%%%%%%%%%%%%%%%%%
\textit{Extended QFI for Lindbladian evolutions.---}Suppose that the dynamics of an open quantum system has the dynamical semigroup property, which has been proven to give rise to the Lindbladian master equation \eqref{open-system}, where the parameter to be estimated is in $\{\mathrm{x}_0,\mathrm{x}_a\}_{a=1}^{k}$. We can replace $H$ with $H-\ave{H}_{\varrho}$ in the dynamical equation with no adverse effect. In the case that all the operators of the set $\{H,\Gamma_{a}\}_{a=1}^k$ commute with each other (as in the dephasing dynamics of example $2$ below), one can obtain
\begin{align}
\partial_{\mathrm{x}_0}\varrho&=-i\mrm \tau[H-\ave{H}_{\varrho},\varrho],
\label{erho-x0}\\
\partial_{\mathrm{x}_a}\varrho&=\tau(\Gamma_{a}\varrho \Gamma_{a}^{\dag}-\frac{1}{2}\{\Gamma_{a}^{\dag}\Gamma_{a},\varrho\}).
\end{align}
When $\varrho$ is invertible \cite{solution-nSLD}, straightforward calculations yield
\begin{align}
L_{0}&=-2i\tau(H-\ave{H}_{\varrho}),
\label{L-x0}\\
L_{a}&=\tau(\Gamma_{a}\varrho \Gamma_{a}^{\dag}\varrho^{-1}-\Gamma_{a}^{\dag}\Gamma_{a}),
\label{L-xi}
\end{align}
as possible choices of the nSLD in the estimation of $\mathrm{x}_0$ and $\mathrm{x}_a$s \cite{solution-nSLD}, respectively. For estimating $\mathrm{x}_0$ Eq.~\eqref{L-x0} gives
\begin{equation}
\mathpzc{F}^{(\mathrm{Q})}_{\mathrm{ext},\mathrm{x}_0}=4\tau^2\Delta_{\varrho}^2 H.
\label{F1-unitary}
\end{equation}
This relation shows that, when estimating the Hamiltonian coupling, the effect of the interaction with environment is encapsulated indirectly only through $\varrho$ in the variance of the Hamiltonian. When the system evolves unitarily the result is the same as Eq.~\eqref{F1-unitary} with $\varrho$ replaced with $\varrho_0$. This result is in complete agreement with the known bound on the QFI \cite{Braunstein-Caves:QFI}.

For the estimation of a jump rate $\mathrm{x}_a$, when $\Gamma_{a}$ commutes with $\{H,\Gamma_b\}_{b\neq a}$, from Eq.~\eqref{L-xi} we obtain
\begin{align}\label{F-lindblad-dynamics}
\mathpzc{F}_{\mathrm{ext,\mathrm{x}_a}}^{(\mathrm{Q})}&=\tau^2\Big(\ave{(\Gamma_{a}^{\dag}\Gamma_{a})^2}_{\varrho}-2\ave{\Gamma_{a}^{{\dag}^2}\Gamma_{a}^2}_{\varrho}+\mathrm{Tr}[\varrho^{-1}(\Gamma_{a}\varrho \Gamma_{a}^{\dag})^2]\Big).
\end{align}
Since this relation exhibits a direct dependence on the \textit{dynamical} properties of the system, it can be useful in studying the role of various features of the open system in the precision of a metrology scenario. Additionally, Eq.~\eqref{F-lindblad-dynamics} may hint which initial quantum state is more suitable---giving a lower estimation error. The following example also shows that our framework can give exact or close-to-exact results.

%%%%%%%%%%%%%%%%%%%%%%%%%%%%%%%%%%%%%%%%%%%%%%%%%%%%%%%%%%%%%%%%%%%
\textit{Example 2: Quantum dephasing channel.---}Consider a dephasing quantum channel defined as $\partial_{\tau}\varrho=\mathrm{x}_1(\sigma_z\varrho\sigma_z-\varrho)/2$. Thus $\partial_{\mathrm{x}_1} \varrho=(\tau/2)(\sigma_z\varrho\sigma_z-\varrho)$. For a separable scenario with the initial state $\varrho_0=(\ketbra{+}{+})^{\otimes N}$ [where $\ket{+}=(\ket{0}+\ket{1})/\sqrt{2}$], Eq.~(\ref{F-lindblad-dynamics}) yields $\mathpzc{F}^{(\mathrm{Q})}_{\mathrm{ext},\mathrm{x}_1}=N\tau^2 e^{-2\tau\mathrm{x}_1}/(1-e^{-2\tau\mathrm{x}_1})$, which is significantly close to the exact value $\mathpzc{F}^{(\mathrm{Q})}_{\mathrm{x}_1}=N\tau^2 e^{-2\tau\mathrm{x}}$ \cite{Chin:PRL2012}. If we choose the entangled $\varrho_0=\ketbra{\textsc{ghz}_N}{\textsc{ghz}_N}$, we get $\mathpzc{F}_{\mathrm{ext},\mathrm{x}_1}^{(\mathrm{Q} )}=N^2 \tau^2e^{-2N\tau\mathrm{x}}/(1-e^{-2N\tau\mathrm{x}})$, in comparison with the slightly different exact QFI $\mathpzc{F}^{(\mathrm{Q})}_{\mathrm{x}_1}=N^2 \tau^2 e^{-2N\tau\mathrm{x}}$.

%%%%%%%%%%%%%%%%%%%%%%%%%%%%%%%%%%%%%%%%%%%%%%%%%%%%%%%%%%%%%%%%%%%
\textit{Summary and outlook.---}Here we have proved an extended convexity property for quantum Fisher information. This property implies that quantum Fisher information of a mixture (convex decomposition) comprising of quantum states with some probabilities is bounded by average quantum Fisher information of constituent states plus classical Fisher information attributed to the mixing probabilities. This division of quantum Fisher information to quantum and classical parts has been shown to have physically interesting and important consequences. For example, we supplemented this convexity property with a notion of non-Hermitian symmetric logarithmic derivative to prove that our convexity relation gives rise to exact value of quantum Fisher information in a general open-system/noisy quantum metrology. The non-Hermitian extension has also been shown to have several appealing physical implications on its own. This concept has enabled us to derive general, closed (and simple) upper bounds on quantum Fisher information for open-system scenarios with Lindbladian (or dynamical semigroup) dynamics. An interesting and practically relevant feature of such bounds is that they clearly relate dynamics to the expected precision of an associated quantum metrology scenario. We have also demonstrated that these bounds are exact or close-to-exact in some important physical systems, and have an intuitive geometrical interpretation.

Another opening that our extended convexity property has made possible is to track how in a quantum metrology scenario precision exhibits a classical (shot-noise) or quantum (sub-shot-noise or Heisenberg) behavior. In particular, we have shown that as the number of probes increases, a competition between classical and quantum parts of Fisher information could determine whether and when (in terms of probe size) to expect either of classical or quantum regimes. It is evident that this possibility can have numerous implications for classical/quantum control and for optimizing a metrology scenario for high-precision advanced technologies in physical (and even biological) systems \cite{experiments}.
%%%%%%%%%%%%%%%%%%%%%%%%%%%%%%%%%%%%%%%%%%%%%%%%%%%%%%%%%%%%%%%%%%%%%

%%%%%%%%%%%%%%%%%%%%%%%%%%%%%%%%%%%%%%%%%%%%%%%%%%%%%%%%%%%%%%%%%%%
\newpage
\appendix
\begin{center}\textbf{Supplemental Material}\end{center}
\section{I. The proof that $\mathpzc{F}^{(\mathrm{Q} )}_{\mathrm{conv}}$  is an upper bound on the QFI of a mixture given by $\varrho=\sum_{a} p_a  \varrho_{a}$}
\label{app:sec1}

The Fisher information of a probabilistic classical system is given by
\begin{equation}\label{classical Fisher information}
\mathpzc{F}_{\mathrm{x}}^{(\mathrm{C})}(\{p\})=\int_{\mathpzc{D}_{\mathrm{x}'}} d\mathrm{x}' \left(\frac{\partial_{\mathrm{x}}p(\mathrm{x}'|\mathrm{x})}{\sqrt{p(\mathrm{x}'|\mathrm{x})}}\right)^2,
\end{equation}
where $p(\mathrm{x}'|\mathrm{x})$ is the conditional probability of obtaining $\mathrm{x}'$ given that $\mathrm{x}$ is the true value of the parameter to be estimated. The quantum version of Fisher information, is derived by inserting $p(\mathrm{x}'|\mathrm{x})=\mathrm{Tr}[\Pi_{\mathrm{x}'}\varrho ]$ into Eq.~(\ref{classical Fisher information}), where $\{\Pi_{\mathrm{x}'}\}$ is a set of measurement operators.
% one can define a derivative operator, $\widetilde{L}_{\varrho}$, as
%\begin{equation}\label{SLD-1}
%\partial_\mathrm{x}\varrho =(\widetilde{L}_{\varrho}\varrho +\varrho  \widetilde{L}_{\varrho}^{\dag})/2,
%\end{equation}
%in which $\widetilde{L}_{\varrho}$ is not necessarily Hermitian. In case $\widetilde{L}_{\varrho}=\widetilde{L}^{\dag}_{\varrho}$, the definition above is equivalent to the definition of SLD.
%
%Inserting $\partial_{\mathrm{x}}\varrho $ from Eq.~(\ref{SLD-1}) and $p(\mathrm{x}'|\mathrm{x})=\mathrm{Tr}[\Pi_{\mathrm{x}'}\varrho ]$ into Eq.~(\ref{classical Fisher information})
%one finds that
%\begin{equation}\label{FI-2}
%\mathpzc{F}_{\mathrm{x}}=\frac{1}{4}\int_{\mathpzc{D}_{\mathrm{x}'}} d\mathrm{x}' \left(\frac{\mrm {Tr}[\Pi_{\mathrm{x}'}\widetilde{L}_{\varrho}\varrho ]+\mrm {Tr}[\Pi_{\mathrm{x}'}\varrho  \widetilde{L}^{\dag}_{\varrho}]}{\sqrt{\mrm {Tr}[\Pi_{\mathrm{x}'}\varrho ]}}\right)^2.
%\end{equation}
%%%%%%%%%%%%%%%%%%%%%%%%%%%%%%%%%%%%%%%%%%%%%%%%%%%%%%%%%%%%%
To prove that $\mathpzc{F}^{(\mathrm{Q} )}_{\mathrm{conv}}$, given in Eq.~(2) [or equivalently Eq.~(4)] of the main text, is an upper bound on the QFI, we differentiate $\varrho=\sum_{a} p_a \varrho_{a}$ through the relation
\begin{equation}
(\textsf{L}_{a} \widetilde{\varrho}_{a}+\widetilde{\varrho}_{a}\textsf{L}_{a})/2=\partial_{\mathrm{x}}\widetilde{\varrho}_{a},
\label{SLDprho}
\end{equation}
in which $\widetilde{\varrho}_{a}=p_a  \varrho_{a}$. Thus, it is obtained that $\partial_{\mathrm{x}}\varrho=\sum_{a}(\textsf{L}_{a} \widetilde{\varrho}_{a}+\widetilde{\varrho}_{a}\textsf{L}_{a})/2$. Using this relation in $\partial_{\mathrm{x}}p(\mathrm{x}'|\mathrm{x})=\mathrm{Tr}[\Pi_{\mathrm{x}'}\partial_{\mathrm{x}}\varrho]$, one obtains from Eq.~\eqref{classical Fisher information} that
%Replacing $p(\mathrm{x}'|\mathrm{x})=\mathrm{Tr}[\Pi_{\mathrm{x}'}\varrho]$ and employing Eq.~(\ref{e-mixture-L}) and $\mrm {Tr}[\Pi_{\mathrm{x}'}\widetilde{L}_{a}\varrho_{a}]^{\ast}=\mrm {Tr}[\Pi_{\mathrm{x}'}\varrho_{a} \widetilde{L}_{a}]$ in Eq.~(3) of the main text give
\begin{align}
\mathpzc{F}_{\mathrm{x}}^{(\mathrm{C})}=~~~~~~~~~~~&\int_{\mathpzc{D}_{\mathrm{x}'}} d\mathrm{x}' \left(\text{Re} \frac{\textstyle{\sum_{a}}\mrm {Tr}[\Pi_{\mathrm{x}'}\mathsf{L}_{a}\widetilde{\varrho}_{a}]}{\sqrt{\mrm {Tr}[\Pi_{\mathrm{x}'}\varrho ]}}\right)^2\nonumber\\
\overset{~~~~~~~~~~~~~~~~~~~~~~~~~~}\leqslant &\int_{\mathpzc{D}_{\mathrm{x}'}} d\mathrm{x}' \left|\frac{\textstyle{\sum_{a}}\mrm {Tr}[\Pi_{\mathrm{x}'}\mathsf{L}_{a}\widetilde{\varrho}_{a}]}{\sqrt{\mrm {Tr}[\Pi_{\mathrm{x}'}\varrho ]}}\right|^2\nonumber\\
\overset{~~~~~~\text{triangle inequality}~~~~~}\leqslant &\int_{\mathpzc{D}_{\mathrm{x}'}} d\mathrm{x}'\sum_{a}\left|\frac{\mrm {Tr}[\Pi_{\mathrm{x}'}\mathsf{L}_{a}\widetilde{\varrho}_{a}]}{\sqrt{\mrm {Tr}[\Pi_{\mathrm{x}'}\varrho ]}}\right|^2\nonumber\\
\overset{\text{Cauchy-Schwarz inequality}}{\leqslant} &\int_{\mathpzc{D}_{\mathrm{x}'}} d\mathrm{x}' \sum_{a}\frac{\mathrm{Tr}[\Pi_{\mathrm{x}'}\widetilde{\varrho}_{a}]}{\mathrm{Tr}[\Pi_{\mathrm{x}'}\varrho]}\mathrm{Tr}[\Pi_{\mathrm{x}'} \mathsf{L}_{a} \widetilde{\varrho}_{a} \mathsf{L}_{a}].
\label{eq16}
\end{align}
%where in the last step we used the Cauchy-Schwarz inequality and the relation $\int_{\mathpzc{D}_{\mathrm{x}'}} d\mathrm{x}'\Pi_{\mathrm{x}'}=\openone$.
To find an upper bound independent of the measurement, we employ the Cauchy-Schwarz inequality, whereby
\begin{equation}
\mathpzc{F}_{\mathrm{x}}^{(\mathrm{C})}\leqslant \int_{\mathpzc{D}_{\mathrm{x}'}} d\mathrm{x}' \sqrt{\frac{\sum_{a} p_a ^2\mathrm{Tr}[\Pi_{\mathrm{x}'}\varrho_{a}]^2}{\mathrm{Tr}[\Pi_{\mathrm{x}'}\varrho]^2}}\sqrt{\sum_{a} p_a ^2\mathrm{Tr}[\Pi_{\mathrm{x}'} \textsf{L}_{a}\varrho_{a} \textsf{L}_{a}]^2}.
\label{eq}
\end{equation}
Since the first term in the right-hand side of Eq.~\eqref{eq} is evidently $\leqslant1$, thus
\begin{align}
\mathpzc{F}_{\mathrm{x}}^{(\mathrm{C})}
&\leqslant \int_{\mathpzc{D}_{\mathrm{x}'}} d\mathrm{x}'\sum_{a}\sqrt{p_a ^2 \mathrm{Tr}\left[\Pi_{\mathrm{x}'} \textsf{L}_{a} \varrho_{a} \textsf{L}_{a}\right]^2}\nonumber\\
&=\sum_{a} p_a \mathrm{Tr}[\varrho_{a} \textsf{L}_{a}^2].
\label{upp-mixture}
\end{align}
From Eq.~\eqref{SLDprho} one can simply find that $\textsf{L}_{a}=\partial_{\mathrm{x}}p_a /p_a +L_{\varrho_a}$. Inserting this relation into the above equation yields
\begin{align}
\mathpzc{F}_{\mathrm{x}}^{(\mathrm{C})}
&\leqslant \mathpzc{F}^{(\mathrm{C})}_{\mathrm{x}}(\{p_a \})+\sum_a p_a\mathpzc{F}^{(\mathrm{Q})}_{\mathrm{x}}(\varrho_a)=:\mathpzc{F}^{(\mathrm{Q})}_{\mathrm{conv}}.
\end{align}

\textit{Remark 1.---}Here we show that any upper bound $\mathpzc{B}_{\mathrm{x}}$ on the classical Fisher information which is \textit{independent} of the chosen family of measurement operators $\mathpzc{P}=\{\Pi_{\mathrm{x}'}\}$ is an upper bound on the QFI too.

\textit{Proof.---}To prove this statement, we note that the classical Fisher information not only depends on $\mathrm{x}$, but also depends on the chosen family of measurement operators $\mathpzc{P}=\{\Pi_{\mathrm{x}'}\}$ [Eq.~\eqref{FI-2}]. Here we indicate this dependence explicitly through the notation $\mathpzc{F}_{\mathrm{x}}^{(\mathrm{C})}(\mathpzc{P})$. Thus, for any $\mathpzc{P}$
%On the one hand, Eq.~\eqref{F-Fe} reads
\begin{equation}
\mathpzc{F}^{(\mathrm{C})}_{\mathrm{x}}(\mathpzc{P})\leqslant \mathpzc{B}_{\mathrm{x}},
\label{F(x,p)}
\end{equation}
in which $\mathpzc{B}_{\mathrm{x}}$ does not depend on $\mathpzc{P}$.
%On the other hand, it has been proved in
%We also know from the literature \cite{Paris:QEforQT} that
%one can find a family of measurement operators, say $\mathpzc{P}^{\star}$, such that
Moreover, since by definition \cite{Paris:QEforQT}
\begin{equation}
\mathpzc{F}_{\mathrm{x}}^{(\mathrm{C})}(\mathpzc{P})\leqslant \max_{\mathpzc{P}}\mathpzc{F}_{\mathrm{x}}^{(\mathrm{C})}(\mathpzc{P})=:\mathpzc{F}^{(\mathrm{Q})}_{\mathrm{x}},
\end{equation}
it is obvious from Eq.~\eqref{F(x,p)} that
\begin{equation}
\max_{\mathpzc{P}}\mathpzc{F}_{\mathrm{x}}^{(\mathrm{C})}(\mathpzc{P})=\mathpzc{F}^{(\mathrm{Q})}_{\mathrm{x}}\leqslant \mathpzc{B}_{\mathrm{x}}.
\end{equation}
\hfill$\blacksquare$\\
Replacing $\mathpzc{B}_{\mathrm{x}}$ with $\mathpzc{F}^{(\mathrm{Q})}_{\mathrm{conv},\mathrm{x}}$, one reaches Eq.~(2) of the main text, i.e.,
\begin{equation}
\mathpzc{F}^{(\mathrm{Q})}_{\mathrm{x}}\leqslant\mathpzc{F}_{\mathrm{conv},\mathrm{x}}^{(\mathrm{Q})}.
\end{equation}
%%%%%%%%%%%%%%%%%%%%%%%%%%%%%%%%%%%%%%%%%%%%%%%%%%%%%%%%%%%%%%%%%%%%%%%%%%%%%%%%%%
\section{II. The proof that $\mathpzc{F}^{(\mathrm{Q})}_{\mathrm{ext}}$ is an upper bound on QFI}\label{app}

%The Fisher information of a probabilistic classical system is given by
%\begin{equation}\label{classical Fisher information}
%\mathpzc{F}_{\mathrm{x}}=\int_{\mathpzc{D}_{\mathrm{x}'}} d\mathrm{x}' \left(\frac{\partial_{\mathrm{x}}p(\mathrm{x}'|\mathrm{x})}{\sqrt{p(\mathrm{x}'|\mathrm{x})}}\right)^2,
%\end{equation}
%where $p(\mathrm{x}'|\mathrm{x})$ is the probability distribution of obtaining $\mathrm{x}'$ from the measurement when $\mathrm{x}$ is the true value of the parameter to be estimated.
%To rewrite the quantum version of Fisher information, one can
Using the definition of the nSLD $\widetilde{L}_{\varrho}$ as an operator satisfying the relation
\begin{equation}\label{SLD-1}
\partial_\mathrm{x}\varrho =(\widetilde{L}_{\varrho}\varrho +\varrho  \widetilde{L}_{\varrho}^{\dag})/2,
\end{equation}
%in which $\widetilde{L}_{\varrho}$ is not necessarily Hermitian. In case $\widetilde{L}_{\varrho}=\widetilde{L}^{\dag}_{\varrho}$, the definition above is equivalent to the definition of SLD.
and inserting $\partial_{\mathrm{x}}\varrho $ from Eq.~(\ref{SLD-1}) and $p(\mathrm{x}'|\mathrm{x})=\mathrm{Tr}[\Pi_{\mathrm{x}'}\varrho ]$ into Eq.~(\ref{classical Fisher information}), one obtains
\begin{equation}\label{FI-2}
\mathpzc{F}_{\mathrm{x}}^{(\mathrm{C})}=\frac{1}{4}\int_{\mathpzc{D}_{\mathrm{x}'}} d\mathrm{x}' \left(\frac{\mrm {Tr}[\Pi_{\mathrm{x}'}\widetilde{L}_{\varrho}\varrho ]+\mrm {Tr}[\Pi_{\mathrm{x}'}\varrho  \widetilde{L}^{\dag}_{\varrho}]}{\sqrt{\mrm {Tr}[\Pi_{\mathrm{x}'}\varrho ]}}\right)^2.
\end{equation}
Since
\begin{eqnarray}
&\mrm {Tr}[\Pi_{\mathrm{x}'}\widetilde{L}_{\varrho}\varrho ]^{\ast}=\mrm {Tr}[\Pi_{\mathrm{x}'}\varrho  \widetilde{L}^{\dag}_{\varrho}],\nonumber
\end{eqnarray}
the Fisher information of Eq.~(\ref{FI-2}) can be rewritten as
\begin{eqnarray}
\mathpzc{F}_{\mathrm{x}}^{(\mathrm{C})}=\frac{1}{2}\int_{\mathpzc{D}_{\mathrm{x}'}} d\mathrm{x}' \left(\left|\frac{\mrm {Tr}[\Pi_{\mathrm{x}'}\widetilde{L}_{\varrho}\varrho ]}{\sqrt{\mrm {Tr}[\Pi_{\mathrm{x}'}\varrho ]}}\right|^2+\frac{(\mathpzc{R}^2-\mathpzc{I}^2)}{\mrm {Tr}[\Pi_{\mathrm{x}'}\varrho ]}\right),
\end{eqnarray}
where by $\mathpzc{R}$ and $\mathpzc{I}$ we mean, respectively, $\mathrm{Re}(\mrm {Tr}[\Pi_{\mathrm{x}'}\widetilde{L}_{\varrho}\varrho ]$ and $\mathrm{Im}(\mrm {Tr}[\Pi_{\mathrm{x}'}\widetilde{L}_{\varrho}\varrho ]$.
Using the Cauchy-Schwarz inequality for the two operators $\sqrt{\Pi_{\mathrm{x}'}}\widetilde{L}_{\varrho}\sqrt{\varrho(\mathrm{x}')}$ and $\sqrt{\varrho_{\mathrm{x}'}}\sqrt{\Pi_{\mathrm{x}'}}$, in the first term, it is found that
\begin{eqnarray}
\mathpzc{F}_{\mathrm{x}}^{(\mathrm{C})}&\leqslant& \frac{1}{2}\int_{\mathpzc{D}_{\mathrm{x}'}} d\mathrm{x}' \left(\mathrm{Tr}[\Pi_{\mathrm{x}'} \widetilde{L}_{\varrho} \varrho  \widetilde{L}_{\varrho}^{\dag}]+\frac{(\mathpzc{R}^2-\mathpzc{I}^2)}{\mrm {Tr}[\Pi_{\mathrm{x}'}\varrho ]}\right)\nonumber\\
&=&\frac{1}{2}\mathrm{Tr}[\widetilde{L}_{\varrho} \varrho \widetilde{L}_{\varrho}^{\dag}]+\frac{1}{2}\int_{\mathpzc{D}_{\mathrm{x}'}} d\mathrm{x}' \frac{(\mathpzc{R}^2-\mathpzc{I}^2)}{\mrm {Tr}[\Pi_{\mathrm{x}'}\varrho ]},
\end{eqnarray}
where we have used $\int_{\mathpzc{D}_{\mathrm{x}'}} d\mathrm{x}'\Pi_{\mathrm{x}'}=\openone$. To find an upper bound independent of the measurement, one can use the following inequality:
\begin{equation}
\frac{(\mathpzc{R}^2-\mathpzc{I}^2)}{\mrm {Tr}[\Pi_{\mathrm{x}'}\varrho ]}\leqslant \frac{(\mathpzc{R}^2+\mathpzc{I}^2)}{\mrm {Tr}[\Pi_{\mathrm{x}'}\varrho ]}=\left|\frac{\mrm {Tr}[\Pi_{\mathrm{x}'}\widetilde{L}_{\varrho}\varrho ]}{\sqrt{\mrm {Tr}[\Pi_{\mathrm{x}'}\varrho ]}}\right|^2,
\end{equation}
and then
\begin{eqnarray}
\mathpzc{F}_{\mathrm{x}}^{(\mathrm{C})}
%\frac{1}{2}\mathrm{Tr}[\widetilde{L}_{\varrho} \varrho  \widetilde{L}_{\varrho}^{\dag}]-\frac{1}{2M}|\mathrm{Tr}[\widetilde{L}_{\varrho}\varrho ]|^2,\nonumber\\
\leqslant \mathrm{Tr}[\widetilde{L}_{\varrho} \varrho  \widetilde{L}_{\varrho}^{\dag}]=: \mathpzc{F}_{\mathrm{ext}, \mathrm{x}}^{(\mathrm{Q})}.
\label{F-Fe}
\end{eqnarray}
%Although this bound may not be achievable, it provides an upper bound for the quantum Fisher information (QFI) in the sense that the QFI is an achievable upper bound on the Fisher information:
%response:

This relation, in which the left-hand side depends on the chosen measurements, is valid for any set of POVMs, in particular the one that maximizes the left hand side to make it equal to the QFI. Hence,
\begin{equation}
\mathpzc{F}^{(\mathrm{Q} )}_{\mathrm{x}}\leqslant\mathpzc{F}^{(\mathrm{Q} )}_{\mathrm{ext},\mathrm{x}}.
\label{FQ-Fe}
\end{equation}
For a detailed reasoning, see the remark in Sec.~\ref{app:sec1}. We also prove Eq.~\eqref{FQ-Fe} in the next section through another approach by directly minimizing the extended QFI.

%%%%%%%%%%%%%%%%%%%%%%%%%%%%%%%%%%%%%%%%%%%%%%%%%%%%%%%%%%%%%%%%%%%%%%%%%
\section{III. Minimization of the extended QFI}

Here we show that minimization of $\mathpzc{F}_{\mathrm{ext}}^{(\mathrm{Q})}$ over different choices of the nSLD has a geometric interpretation as the ``Uhlmann parallel transport condition," which leads to the QFI as the minimum of the extended QFI.

Any quantum state $\varrho$ can be decomposed as $\varrho=\textit{w}\textit{w}^{\dag}$, where $\textit{w}=\sqrt{\varrho}U$, with $U$ an arbitrary ($\mathrm{x}$-dependent) unitary operator. Thus, we have $\partial_{\mathrm{x}}{\varrho}=(\partial_{\mathrm{x}}  \textit{w})\textit{w}^{\dag}+  \textit{w}(\partial_{\mathrm{x}}\textit{w}^{\dag})$. Comparing this relation with Eq.~(8) in the main text shows that a consistent choice for an nSLD is
\begin{equation}
\widetilde{L}_{\varrho}=2(\partial_{\mathrm{x}}\textit{w})\textit{w}^{-1}.
\label{nSLD}
\end{equation}
Hence, for every quantum state, the extended QFI becomes
\begin{eqnarray}\label{QFI-1}
\mathpzc{F}^{(\mathrm Q)}_{\mathrm{ext},\mathrm{x}}=4~\mathrm{Tr}[(\partial_{\mathrm{x}} \textit{w})(\partial_{\mathrm{x}}\textit{w}^{\dag})].
\end{eqnarray}
This form is reminiscent (except the prefactor $4$) of the very Uhlmann metric \cite{Uhlmann:PT}, which has been shown to be minimized when the following parallel transport condition
\begin{equation}
\textit{w}^{\dag}\partial_{\mathrm{x}}\textit{w}=(\partial_{\mathrm{x}} \textit{w}^{\dag}) \textit{w}
\end{equation}
is satisfied. If we rewrite this equation in terms of the nSLD of Eq.~\eqref{nSLD},
it is obtained that the parallel transport condition leads to the Hermiticity of the nSLD,
\begin{eqnarray}
\textit{w}^{\dag}\partial_{\mathrm{x}}\textit{w}&=&\left(\partial_{\mathrm{x}} \textit{w}^{\dag}\right)\textit{w}\nonumber\\
\textit{w}^{\dag}\big(\partial_{\mathrm{x}}\textit{w}\big)\textit{w}^{-1}&=&\partial_{\mathrm{x}} \textit{w}^{\dag}\nonumber\\
\big(\partial_{\mathrm{x}}\textit{w}\big) \textit{w}^{-1}&=&\textit{w}^{\dag^{-1}}\partial_{\mathrm{x}} \textit{w}^{\dag}\nonumber\\
2\big(\partial_{\mathrm{x}}\textit{w}\big)\textit{w}^{-1}&=&2[\big(\partial_{\mathrm{x}} \textit{w}\big)\textit{w}^{-1}]^{\dag}\nonumber\\
\widetilde{L}_{\varrho}=\widetilde{L}^{\dag}_{\varrho}.
\end{eqnarray}
Thus, when the parallel transport condition is satisfied, the extended QFI is equivalent to the QFI.

Equation \eqref{QFI-1} is not $U$-invariant, i.e., choosing different purifications with a parameter-dependent unitary leads to different values for $\mathpzc{F}^{(\mathrm Q)}_{\mathrm{ext}}$ as
%Employing the parallel transport condition on Eq.~\eqref{QFI-1} leads to
\begin{eqnarray}
\mathpzc{F}^{(\mathrm Q)}_{\mathrm{ext},\mathrm{x}}=4~\mathrm{Tr}[(\partial_{\mathrm{x}}\sqrt{\varrho})^2+\partial_{\mathrm{x}}U^{\dag}\varrho \partial_{\mathrm{x}}U)].
\end{eqnarray}
The minimum upper bound for the QFI is obtained when $U$ is the unitary satisfying the parallel transport condition. As a special case, if an $\mathrm{x}$-independent $U$ is chosen, we obtain
\begin{eqnarray}
\label{QFI-2}
\mathpzc{F}^{(\mathrm Q)}_{\mathrm{ext},\mathrm{x}}=4~\mathrm{Tr}[(\partial_{\mathrm{x}}\sqrt{\varrho})^2].
\end{eqnarray}
Using this simplified relation for a pure state $\varrho=|\psi\rangle\langle \psi|$, since $\sqrt{\varrho}=\varrho$, the following simplification emerges:
\begin{equation}
\mathpzc{F}^{(\mathrm Q)}_{\mathrm{ext},\mathrm{x}}=8 \left(\braket{\partial_{\mathrm{x}}\psi}{\partial_{\mathrm{x}}\psi}-\left|\braket{\partial_{\mathrm{x}}\psi}{\psi}\right|^2\right),
\end{equation}
which is twice QFI ($\mathpzc{F}^{(\mathrm{Q})}_{\mathrm{x}}$). For the case of a mixed quantum state $\varrho$ which evolves with the Hamiltonian $\mathrm{x} H$, we have $\partial_{\mathrm{x}}\sqrt{\varrho}=-i \tau[H,\sqrt{\varrho}]$, thus Eq.~(\ref{QFI-2}) yields $\mathpzc{F}^{(\mathrm Q)}_{\mathrm{ext},\mathrm{x}}=-4\tau^2~\mathrm{Tr}\left[[H,\sqrt{\varrho}]^2\right]$, which is $8 \tau^2$ times the skew information \cite{Skew-info}.

%Restating the parallel transport condition in terms of nSLD shows that $\widetilde{L}_{\varrho}$ must be Hermitian \cite{supp}, the case for which %$\mathpzc{F}^{(\mathrm{Q})}_{\mathrm{ext}}$ would be the exact $\mathpzc{F}^{(\mathrm{Q})}$.

\textit{Remark 2.---}It should be noted that Eq.~(\ref{QFI-2}) is another natural quantization of the classical Fisher information which will be proved in the next section. %\ref{rem2}.

%%%%%%%%%%%%%%%%%%%%%%%%%%%%%%%%%%%%%%%%%%%%%%%%%%%%%%%%%%%%%%%%%%%%%%%
\section{IV. Natural quantizations of the classical Fisher information as upper bounds on the QFI}\label{SLD-independent}
\begin{itemize}
\item \label{rem2}
If one rewrites the classical Eq.~(3) as $\mathpzc{F}_{\mathrm{x}}^{(\mathrm{C})}=4\int_{\mathpzc{D}_{\mathrm{x}'}}d\mathrm{x}'\Big(\partial_{\mathrm{x}}\sqrt{p(\mathrm{x}'|\mathrm{x})}\Big)^2$, and next replaces $p(\mathrm{x}'|\mathrm{x})$ with $\varrho$ and the integration over $\mathrm{x}'$ with trace, a quantization of the classical Fisher information as Eq.~(17) is obtained. In fact, this is the very approach through which the quantity defined in  Eq.~(7) had already been appeared in the literature \cite{Skew-info}, but without noting that this quantity is an upper bound on the QFI.

\item
Writing Eq.~(3) as $\mathpzc{F}^{(\mathrm{C})}_{\mathrm{x}}=\int_{\mathpzc{D}_{\mathrm{x}'}}d\mathrm{x}'(\partial_{\mathrm{x}}p(\mathrm{x}'|\mathrm{x}))^2/p(\mathrm{x}'|\mathrm{x})$ and then, the same as above, replacing $p(\mathrm{x}'|\mathrm{x})$ with $\varrho$ and the integration over $\mathrm{x}'$ with trace, another natural quantization of the classical Fisher information is obtained as $\mathrm{Tr}[\varrho^{-1}\left(\partial_{\mathrm{x}}\varrho\right)^2]$.

\textit{Proof. }Since $\mathrm{Tr}[\Pi_{\mathrm{x}'}(\widetilde{L}_{\varrho}\varrho +\varrho  \widetilde{L}^{\dag}_{\varrho})]=\mathrm{Tr}[\Pi_{\mathrm{x}'}\varrho (\varrho^{-1}\widetilde{L}_{\varrho}\varrho +\widetilde{L}^{\dag}_{\varrho})]$. Following Eq.~(\ref{FI-2}) and using the Cauchy-Schwarz inequality, it is found that
\begin{widetext}
\begin{eqnarray}
\mathpzc{F}^{(\mathrm{C})}_{\mathrm{x}}&&=\frac{1}{4}\int_{\mathpzc{D}_{\mathrm{x}'}} d\mathrm{x}'\Big|\mathrm{Tr}\Big[\frac{\sqrt{\Pi_{\mathrm{x}'}}\sqrt{\varrho }}{\sqrt{\mathrm{Tr}[\Pi_{\mathrm{x}'}\varrho ]}}
\sqrt{\varrho }\left(\varrho^{-1}\widetilde{L}_{\varrho}\varrho +\widetilde{L}^{\dag}_{\varrho}\right)\sqrt{\Pi_{\mathrm{x}'}}\Big]\Big|^2\nonumber\\
&&\leqslant \frac{1}{4}\int_{\mathpzc{D}_{\mathrm{x}'}} d\mathrm{x}'\mathrm{Tr}[\varrho \left(\varrho^{-1}\widetilde{L}_{\varrho}\varrho +\widetilde{L}^{\dag}_{\varrho}\right)\Pi_{\mathrm{x}'}\left(\varrho \widetilde{L}^{\dag}_{\varrho}\varrho^{-1}+\widetilde{L}_{\varrho}\right)]\nonumber\\
&&=\frac{1}{4}\mathrm{Tr}\left[\varrho \left(\varrho^{-1}\widetilde{L}_{\varrho}\varrho +\widetilde{L}^{\dag}_{\varrho}\right)\left(\varrho \widetilde{L}^{\dag}_{\varrho}\varrho^{-1}+\widetilde{L}_{\varrho}\right)\right]\nonumber\\
&&=\frac{1}4\mathrm{Tr}\left[\widetilde{L}_{\varrho}\varrho^2 \widetilde{L}^{\dag}_{\varrho}\varrho^{-1} +\widetilde{L}_{\varrho}\varrho \widetilde{L}_{\varrho}+\widetilde{L}^{\dag}_{\varrho}\varrho \widetilde{L}^{\dag}_{\varrho}+\widetilde{L}_{\varrho}\varrho \widetilde{L}^{\dag}_{\varrho}\right]\nonumber\\
&&=\frac{1}4\mathrm{Tr}\left[(\varrho^{-1}\widetilde{L}_{\varrho}\varrho +\widetilde{L}^{\dag}_{\varrho})\varrho \widetilde{L}^{\dag}_{\varrho}+\widetilde{L}_{\varrho}\left(\widetilde{L}_{\varrho}\varrho +\varrho \widetilde{L}^{\dag}_{\varrho}\right)\right]\nonumber\\
&&=\frac{1}4\mathrm{Tr}\left[\varrho \widetilde{L}^{\dag}_{\varrho}\varrho^{-1}(\widetilde{L}_{\varrho}\varrho +\varrho \widetilde{L}^{\dag}_{\varrho})+\widetilde{L}_{\varrho}\left(\widetilde{L}_{\varrho}\varrho +\varrho \widetilde{L}^{\dag}_{\varrho}\right)\right]\nonumber\\
&&=\frac{1}4\mathrm{Tr}\left[\left(\varrho \widetilde{L}^{\dag}_{\varrho}\varrho^{-1}+\widetilde{L}_{\varrho}\right)\left(\widetilde{L}_{\varrho}\varrho +\varrho \widetilde{L}^{\dag}_{\varrho}\right)\right]\nonumber\\
&&=\frac{1}4\mathrm{Tr}\left[\left(\varrho \widetilde{L}^{\dag}_{\varrho}+\widetilde{L}_{\varrho}\varrho \right)\varrho^{-1}\left(\widetilde{L}_{\varrho}\varrho +\varrho \widetilde{L}^{\dag}_{\varrho}\right)\right]\nonumber\\
&&=\frac{1}4\mathrm{Tr}\left[\varrho^{-1}\left(\widetilde{L}_{\varrho}\varrho +\varrho \widetilde{L}^{\dag}_{\varrho}\right)^2\right]\nonumber\\
&&=\mathrm{Tr}\left[\varrho^{-1}(\partial_{\mathrm{x}}\varrho )^2\right].
\end{eqnarray}
\end{widetext}
The bound is saturated if $\{\Pi_{\mathrm{x}'}\}$ is chosen such that
\begin{eqnarray}
\frac{\sqrt{\varrho }\sqrt{\Pi_{\mathrm{x}'}}}{\mathrm{Tr}[\Pi_{\mathrm{x}'}\varrho ]}=\frac{\sqrt{\varrho }(\varrho^{-1}\widetilde{L}_{\varrho}\varrho +\widetilde{L}^{\dag}_{\varrho})\sqrt{\Pi_{\mathrm{x}'}}}{\mathrm{Tr}[\varrho  (\varrho^{-1}\widetilde{L}_{\varrho}\varrho +\widetilde{L}^{\dag}_{\varrho}) \Pi_{\mathrm{x}'}]},\nonumber\\
\end{eqnarray}
which in turn is satisfied by choosing the eigenvectors of $\varrho^{-1}\widetilde{L}_{\varrho}\varrho +\widetilde{L}^{\dag}_{\varrho}$ as $\{\Pi_{\mathrm{x}'}\}$. However, since $\varrho^{-1}\widetilde{L}_{\varrho}\varrho +\widetilde{L}^{\dag}_{\varrho}$ is not necessarily Hermitian, its eigenvectors do not provide a complete set for measurement operators. As a result, the bound is not necessarily achievable.
\end{itemize}
%%%%%%%%%%%%%%%%%%%%%%%%%%%%%%%%%%%%%%%%%%%%%%%%%%%%%%%%%%%%%%%%%%%%%%%%%%%%%%%%%
%%%%%%%%%%%%%%%%%%%%%%%%%%%%%%%%%%%%%%%%%%%%%%%%%%%%%%%%%%%%%%%%%%%%%%%%%%%%%%%%%
%\section{F. Proof of EQ.~(16)} %eqref{eq16}
%\begin{align}
%\mathpzc{F}_{\mathrm{x}}&=\int_{\mathpzc{D}_{\mathrm{x}'}} d\mathrm{x}' \left(\text{Re} \frac{\textstyle{\sum_{a}}\mrm {Tr}[\Pi_{\mathrm{x}'}\widetilde{L}_{a}\varrho_{a}]}{\sqrt{\mrm {Tr}[\Pi_{\mathrm{x}'}\varrho ]}}\right)^2\nonumber\\
%&\leqslant \int_{\mathpzc{D}_{\mathrm{x}'}} d\mathrm{x}' \left|\frac{\textstyle{\sum_{a}}\mrm {Tr}[\Pi_{\mathrm{x}'}\widetilde{L}_{a}\varrho_{a}]}{\sqrt{\mrm {Tr}[\Pi_{\mathrm{x}'}\varrho ]}}\right|^2\nonumber\\
%&\leqslant \int_{\mathpzc{D}_{\mathrm{x}'}} d\mathrm{x}'\textstyle{\sum_{a}}\left|\frac{\mrm {Tr}[\Pi_{\mathrm{x}'}\widetilde{L}_{a}\varrho_{a}]}{\sqrt{\mrm {Tr}[\Pi_{\mathrm{x}'}\varrho ]}}\right|^2\nonumber\\
%&\leqslant \int_{\mathpzc{D}_{\mathrm{x}'}} d\mathrm{x}' \textstyle{\sum_{a}}\frac{\mathrm{Tr}[\Pi_{\mathrm{x}'}\varrho_{a}]}{\mathrm{Tr}[\Pi_{\mathrm{x}'}\varrho]}\mathrm{Tr}[\Pi_{\mathrm{x}'} \widetilde{L}_{a} \varrho_{a} \widetilde{L}_{a}],
%\end{align}
%where in the last step we used the Cauchy-Schwarz inequality and the relation $\int_{\mathpzc{D}_{\mathrm{x}'}} d\mathrm{x}'\Pi_{\mathrm{x}'}=\openone$.

%%%%%%%%%%%%%%%%%%%%%%%%%%%%%%%%%%%%%%%%%

\end{document}